\documentstyle[aaspp4]{article}
\lefthead{Mori et al.}
\righthead{Formation of dwarf elliptical galaxies}

\def\r{\mbox{\boldmath{$r$}}}
\def\rj{\mbox{\boldmath{$r$}}_j}
\def\v{\mbox{\boldmath{$v$}}}
\def\g{\mbox{\boldmath{$g$}}}
\def\AA{$\stackrel{\ _\circ}{A}$ }

\begin{document}

\title{
Dissipative Process as a Mechanism of Differentiating
Internal Structures between Dwarf and Normal Elliptical 
Galaxies in a CDM Universe
}

\author{Masao Mori}

\affil{
Max-Planck-Institut f\"{u}r Astronomie,
K\"{o}nigstuhl 17, D-69117 Heidelberg, Germany\\
Institute of Astronomy, University of Tokyo, 
Mitaka, Tokyo 181-8588, Japan
}

\author{Yuzuru Yoshii}
\affil{
Institute of Astronomy, University of Tokyo, 
Mitaka, Tokyo 181-8588, Japan\\
Research Center for the Early Universe, 
University of Tokyo, Bunkyo-ku, Tokyo 113-0033, Japan
}

\and

\author{Ken'ichi Nomoto}
\affil{
Department of Astronomy, University of Tokyo, 
Bunkyo-ku, Tokyo 113-0033, Japan\\
Research Center for the Early Universe, 
University of Tokyo, Bunkyo-ku, Tokyo 113-0033, Japan
}

\received{August 4, 1998}
\accepted{September 4, 1998}

\begin{abstract}
We simulate the dynamical, chemical, and spectro-photometric evolution 
of dwarf and normal elliptical galaxies embedded in a dark matter halo, 
using a three-dimensional $N$-body/SPH simulation code. For the forming 
dwarf elliptical galaxies, supernova-driven winds propagating 
outwards from inside the system collide with the infalling gas and 
produce the super-shell in which stars are formed.   
The resulting stellar system forms a loosely bound virialized system
due to the significant mass loss and has a large velocity dispersion 
and a large core. Consequently the surface brightness distribution shows
an exponential profile and the color distribution shows a {\it positive}
gradient such that the colors become redder away from the galaxy center 
in agreement with observations. 
On the other hand, for the normal elliptical galaxies with deep 
gravitational potential, the mass loss out of the system does not have a 
significant dynamical effect on the stellar system. The resulting surface 
brightness distribution has a large central concentration like 
de Vaucouleurs' $r^{1/4}$-profile and the color distribution shows a
{\it negative} gradient as observed. Our simulation shows that different 
features between dwarf and normal elliptical galaxies stem from different 
cooling efficiencies for their respective protogalaxies in a standard 
CDM universe.
\end{abstract}


\section{INTRODUCTION}
Recent progress in observational devices and techniques has enhanced
our knowledge of formation and evolution of galaxies on a firm statistical
basis. The luminosity function of galaxies observed in general fields and 
in nearby clusters shows a steep slope in both the bright and faint ends,
whereas its slope in between exhibits a plateau or a slight dip from 
$M_{\rm B}=-19$ to $-13$ mag (Loveday 1997; Trentham 1998). 

Hierarchical clustering models of galaxy formation in a standard cold 
dark matter (CDM) universe predict that the number of galaxies 
monotonically increases with decreasing mass, yielding a power-law mass 
function in a low-mass end
(Davis et al. 1985; White \& Frenk 1991; Cole et al. 1994).
This implies that, unless the mass-to-light ratio is very different from 
what we usually expect, an excessive number of dwarf galaxies are predicted 
beyond that observed in the luminosity function.

In order to remove this serious discrepancy several mechanisms for 
suppressing the formation of dwarf galaxies have been proposed so far.
Efstathiou (1992) proposed an effect of photoionization by ultraviolet 
background radiation that keeps the gas hot and unable to collapse 
(see also Chiba \& Nath 1994; Thoul \& Weinberg 1995).
Dekel \& Silk (1986) proposed an energy feedback that induces a significant 
mass loss from low-mass galaxies having the shallow gravitational potential
(see also Saito 1979; Yoshii \& Arimoto 1987; Lacey \& Silk 1991).

A clear distinction is seen in structural and photometric quantities 
between dwarf and normal ellipticals in spite of their morphological 
similarity. 
Faber \& Lin (1983) appreciated that dwarf ellipticals have an exponential 
surface brightness profile rather than de Vaucouleurs' $r^{1/4}$-profile 
applied to normal ellipticals (de Vaucouleurs 1948; Kormendy 1977).
Bright dwarfs with $M_{\rm B}<-16$ mag show a distinct luminosity spike
in their center commonly referred to as the nucleated dwarfs 
(Caldwell \& Bothun 1987; Binggeli \& Cameron 1991; 
Ichikawa, Wakamatsu \& Okamura 1986), and faint dwarfs do not usually show
such a nucleus (Sandage et al. 1985). Young \& Currie (1994) and 
Binggeli \& Jerjen (1998) showed that the surface brightness profile
of dwarf ellipticals changes from an exponential to an $r^{1/4}$-like form 
as their luminosity increases.

Vader et al. (1988) analysed the data from Vigroux et al. (1988)
and found that many dwarf ellipticals show systematically redder colors at 
larger radii away from the galaxy center (Vader et al. 1988; 
Kormendy \& Djorgovski 1989; Chaboyer 1994; Durrell et al. 1996), 
which is opposite to the trend observed in normal ellipticals.
Vader et al. (1988) interpreted this inverse color gradient in
dwarf galaxies in term of the positive gradient of stellar age across the 
system. However, because of their very low metallicities, stars in dwarf 
galaxies must have been formed on very short time scales and therefore no 
appreciable age difference results.

Recent high-resolution observations of nearby dwarf galaxies show a web of 
filaments, loops and expanding super-giant shells which are imprinted in 
the gas around the individual galaxies and such a striking feature likely
stems from the interaction of interstellar medium with energetic stellar 
winds from massive stars and/or supernova explosions
(Meurer, Freeman \& Dopita 1992; Puche \& Westpfahl 1994; 
Westpfahl \& Puche 1994; Marlowe, Heckman \& Wyse 1995; Hunter 1996).
All these observations motivate us to consider that the energy feedback 
from supernovae may be of prime importance to understand the discrepancy 
between the mass function of galaxies predicted from hierarchical models 
and the luminosity function obtained from recent redshift surveys.

From theoretical viewpoints, while collisionless $N$-body simulations
are successful in producing an $r^{1/4}$-profile 
through the merging of galaxies (White 1979) as well as their monolithic 
collapse (van Albada 1982), only a few simulations incorporating the
star formation and energy feedback from supernovae have been performed 
(e.g. Theis, Burkert \& Hensler 1992; Mori et al. 1997) and in most cases 
a simple one-zone chemical evolution model has yet been used to interpret 
the data (e.g. Dekel \& Silk 1986; Yoshii \& Arimoto 1987; 
Babul \& Ferguson 1996).
Navarro, Eke \& Frenk (1996) recently studied the dynamical response of
a virialized system to an impulsive mass loss from the galaxy center based
on collisionless $N$-body simulations. However, their assumption of impulsive 
mass loss needs to be justified because it is not known a priori whether
the time scale of mass loss is shorter than the dynamical time of the system.
Therefore, more realistic simulations are necessary in order to
model the evolution of galaxies and properly interpret the accumulated 
data including the spatial gradients of structural and photometric quantities.

In this paper, using a powerful hybrid code for three-dimensional $N$-body
and hydrodynamical simulations, we present a unified view of the dynamical,
chemical and spectro-photometric evolution of dwarf and normal elliptical
galaxies from cosmologically motivated initial conditions.
In \S 2 we briefly describe the basic equations for the model and the 
method of incorporating the cooling, star formation, and energy feedback 
process in the model. More detailed description of the numerical method is
deferred to the paper by Mori, Nakasato \& Nomoto (1998).
In \S 3 we carry out the simulations for dwarf and normal 
elliptical galaxies originated from a $1\sigma$ density perturbation in 
a CDM universe. In \S 4 we summarise the result of this paper.

\section{MODELS}

Our three-dimensional simulation method combines the collisionless dynamics 
with the hydrodynamics in a consistent manner.
The evolution of a collisionless system consisting of dark matter and stars 
is followed by an $N$-body code, and the hydrodynamical properties 
of a gaseous component are calculated using the {\it Smoothed Particle 
Hydrodynamics} (SPH) (Lucy 1977; Gingold \& Monaghan 1977). 
The SPH equations have a similar structure to the self-gravitational 
$N$-body system. The SPH has no restriction on spatial resolution or 
deviation from any symmetry and provides a surprisingly accurate result 
in some applications even with a small number of particles. 
These characteristics make the SPH most suited to the study of formation
and evolution of galaxies.

\subsection{Basic equations}

The gaseous component is described by the fluid equation for a perfect 
gas in the Lagrangian form. 
The smoothed average for a hydrodynamical quantity $f(\r)$ is given by
\begin{eqnarray}
<f(\r)> = \int f(\r') W(\r-\r',h) d^3\r',
\label{eqn:basic}
\end{eqnarray}
provided $\int W(\r-\r',h) d^3\r' = 1$ and 
$\lim_{h \to 0} W(\r-\r',h) = \delta(\r-\r')$,
where $W(\r,h)$ is the smoothing kernel and $h$ is the smoothing length.
For nearby gas particles, we replace the integration by summation because 
of the finite number of fluid elements.
For example, the gas density $\rho_{\rm g}$ at position $\r$ 
is given by
\begin{eqnarray}
\rho_{\rm g}(\r) = \sum_j m_j W( |\r-\rj|, h ),
\end{eqnarray}
where $m_j$ is the mass of gas particle located at position $\r_j$.
In this paper, following Monaghan \& Lattanzio (1985), we use the standard 
kernel of spherically symmetric spline.

The momentum equation is given by
\begin{eqnarray}
\frac{d\v_{\rm g}}{dt} = -\frac{1}{\rho_{\rm g}} \nabla P + \g,
\end{eqnarray}
and the thermal energy equation associated with the rates of cooling 
$ \Lambda $ and heating $ \Gamma $ is given by
\begin{eqnarray}
\frac{d \varepsilon}{dt} 
= -\frac{P}{\rho_{\rm g}} \nabla \cdot \v_{\rm g}
 + \frac{\Gamma - \Lambda }{\rho_{\rm g}},
\end{eqnarray}
with
\begin{eqnarray}
P = ( \gamma -1 ) \rho_{\rm g} \varepsilon,
\end{eqnarray}
where $\v_{\rm g}$ is the gas velocity, $\g$ is the gravitational force, 
$P$ is the gas pressure, $\gamma$ (=5/3) is the adiabatic index, and 
$\varepsilon$ is the specific internal energy. 
We assume that the gas is optically thin and in ionization equilibrium.
In order to focus mostly on the heating by energy feedback from supernovae 
to the interstellar medium, we neglect the effect of photoionization by 
UV background radiation for simplicity.
We use the radiative cooling taken from Table 6 of Sutherland \& Dopita 
(1993). In Fig. 1 the cooling rate is plotted against the temperature
for different metallicities. The top curve shows the cooling rate for 
[Fe/H]=+0.5, and the lower curves show those with decreasing the metallicity 
at intervals of 0.5 dex.
The bottom curve corresponds to the zero metallicity for the primordial gas 
composition.

The equation of motion for a collisionless particle of either dark matter 
or star at $\r$ is given by
\begin{eqnarray}
\frac{d \v  }{d t} & = & \g,  
\label{eqn:grav1}
\end{eqnarray}
with
\begin{eqnarray}
\g & = & -G \sum_{j}^{N}  
      \frac{m_j}{\{ (\r-\rj)^2 + \epsilon^2 \}^{3/2}}, 
\label{eqn:grav2}
\end{eqnarray}
where $G$ is the gravitational constant and $\epsilon$ is the softening 
parameter.
We calculate the gravitational force $\g$ by using the ``Remote-GRAPE'' 
system (Nakasato, Mori \& Nomoto 1997) for which GRAPE refers to a special 
purpose computer for efficiently calculating the gravitational force and 
potential (Sugimoto et al. 1990).
Self-gravity calculations can be performed in parallel with 
other calculations such as gasdynamics, star formation and feedback, 
so that the total calculation time is considerably shortened. 
The performance analysis is reported by Nakasato, Mori \& Nomoto (1997).

\subsection{Physical processes}

In each time step we calculate at each fluid point three time scales of
the local gas dynamical time ($t_{\rm dyn}$), the local cooling time 
($t_{\rm cool}$), and the sound crossing time ($t_{\rm sound}$):
\begin{eqnarray}
t_{\rm dyn} = \sqrt{\frac{3 \pi }{32 G \rho_{\rm g}}},
\end{eqnarray}
\begin{eqnarray}
t_{\rm cool} = \frac{3}{2} \frac{1}{\mu^2(1-Y)^2} 
                  \frac{k_{\rm B}T}{n_{\rm g} \Lambda},
\end{eqnarray}
and
\begin{eqnarray}
t_{\rm sound} = \frac{l}{c_{\rm s}},
\end{eqnarray}
where $\rho_{\rm g}=\mu m_{\rm H} n_{\rm g}$, $\mu$ is the mean molecular 
weight, $m_{\rm H}$ is the hydrogen mass, $n_{\rm g}$ is the gas number 
density, $T$ is the gas temperature, $Y$ (=0.25) is the helium mass fraction,
$k_{\rm B}$ is the Boltzmann constant, $c_{\rm s}$ is local sound 
speed, and $l$ is the local scale length of the fluid.
Here we set $l$ to be equal to the smoothing length of a gas particle.
In our simulation stars are assumed to form in the rapidly cooling, 
Jeans-unstable, and converging region subject to
\begin{eqnarray}
t_{\rm cool} < t_{\rm dyn} < t_{\rm sound}
{\ \ \ \rm and \ \ \ }
\nabla \cdot \v < 0.
\label{eqn:sfc}
\end{eqnarray}
Once a region simultaneously satisfying these criteria has been identified, 
we create new collisionless {\it star particles} there at a rate determined 
by local gas properties. The subsequent motion of {\it star particles} thus 
formed is determined only by gravity.

The criteria in equation (\ref{eqn:sfc}) are similar to those used by 
Katz (1992), Navarro \& White (1993), and Steinmetz \& M\"uller (1994).
However, they prescribed that about one third or half of the mass of a gas
particle forms a new collisionless {\it star particle} and the rest is heated 
by supernovae. 
In contrast, we do not fix the mass fraction of a gas particle that
is converted to a new {\it star particle}. 
We assume that the star formation rate (SFR) is proportional to the local 
gas density and inversely proportional to the local dynamical time:
\begin{eqnarray}
\frac{d\rho_{\rm s}}{dt} = C \frac{\rho_{\rm g}}{t_{\rm dyn}},
\end{eqnarray}
where the SFR coefficient $C$ is treated as a free parameter according to 
the analysis by Katz, Weinberg \& Hernquist (1995).
We note that our simulation is insensitive to the adopted value of this
parameter (cf. \S 3.1).

We estimate the mass of a newly born {\it star particle} as
\begin{eqnarray}
m_{\rm s} = \left\{
            1-\exp \left( - C \frac{\triangle t}{t_{\rm dyn}} \right)
            \right\} \pi l^3 \rho_{\rm g},
\end{eqnarray}
where $\triangle t$ is the time step used (Mori, Nakasato \& Nomoto 1998). 
Given the slope index $x$ and lower and upper mass limits ($m_{\rm l}, 
m_{\rm u}$) for the initial stellar mass function (IMF), the number of 
Type II supernovae (SNe II) progenitors for a {\it star particle} with 
mass $m_{\rm s}$ is calculated as
\begin{eqnarray}
N_{\rm SN} = \frac{x-1}{x} \ 
         \frac{1-(m_{\rm SN, l}/m_{\rm u})^{-x}}
              {1-(m_{\rm l}/m_{\rm u})^{1-x}} 
	 \frac{m_{\rm s}}{m_{\rm u}},
\label{eqn:sn}
\end{eqnarray}
where $m_{\rm SN, l}=8 M_\odot$ is the lower mass limit of stars that 
will explode as SNe II.
These IMF parameters affect the heating rate of interstellar medium and 
the ejection rate of heavy elements from the {\it star particle}.
The number of SNe II is the most sensitive to the IMF slope among 
others.
Current resolution in our simulation gives $m_{\rm s} \sim 10^{4-6} M_\odot$ 
far exceeding a typical mass of single stars. Therefore, we distribute the 
associated mass of the {\it star particle} over approximately $10^{2-4}$ 
single stars according to Salpeter's (1955) IMF ($x=1.35$) provided that 
the lower and upper mass limits are taken as $m_{\rm l}=0.1M_\odot$ and 
$m_{\rm u}=60M_\odot$, respectively. 

Stars once formed work as sources of transferring the energy, 
synthesized heavy elements, and materials (H and He) to the 
interstellar medium through supernovae or stellar winds from massive stars. 
This feedback process is most critical in the simulation of galaxy 
formation. However, previous authors adopted different treatments, mostly
because there is no good appreciation of how it should be modeled in the
SPH algorithm (Katz 1992; Navarro \& White 1993; Mihos \& Hernquist 1994). 

In this paper, when a {\it star particle} is formed and identified with a 
stellar assemblage as described above, stars more massive than 8 M$_{\odot}$ 
start to explode as SNe II with the explosion energy of $ 10^{51} $ ergs 
and their outer layers are blown out with synthesized heavy elements into 
the interstellar medium leaving the 1.4 M$_\odot$ remnant.
We can regard this stellar assemblage as continuous energy release 
at an average rate of $L_{\rm SN} = 8.44 \times 10^{35}$ ergs s$^{-1}$ 
per star during the explosion period from $\tau(m_{\rm u})=5.4\times 10^6$ 
yr until $\tau(8 M_\odot)=4.3 \times 10^7$ yr,
where $\tau(m)$ is the lifetime of a star of mass $m$ 
(David, Forman \& Jones 1990).
Prior to the onset of SNe II explosions, however, their progenitors 
develop stellar winds and also release the energy of $10^{50}$ ergs 
into the interstellar medium at an average rate of 
$L_{\rm SW} = 7.75 \times 10^{34}$ ergs s$^{-1}$ per star.  
The released energy from stellar winds is supplied to the gas particles 
within a sphere of radius $R_{\rm snr}$, and the energy, heavy elements and 
materials from SNe II are subsequently supplied to the same region.  
The radius $R_{\rm snr}$ is set to be equal to the maximum extension of 
the shock front in the adiabatic phase of supernova remnant and given by 
Shull \& Silk (1979),
\begin{eqnarray}
R_{\rm snr} = 32.9 \left( \frac{ E }{10^{51} \, {\rm ergs}} \right)^{1/4}
          \left( \frac{ n_{\rm g}}{1\, {\rm cm^{-3}}} \right)^{-1/2} {\rm pc},
\end{eqnarray}
where $E$ is the released energy.
In each time step we estimate $n_{\rm g}$ equal to the number density of the 
surrounding gas of the {\it star particle} with the minimum limit of 
$10^{-4}$ cm$^{-3}$.
The gas within $R_{\rm snr}$ remains adiabatic until the multiple SNe II phase 
ends at $\tau(8M_\odot)$, and then it cools according to the adopted cooling
rate of the gas.

Tsujimoto et al. (1996) tabulated the masses of 27 heavy elements synthesized
in the progenitors of SN Ia and II, and determined the number ratio of 
Type Ia supernovae (SNe Ia) relative to SNe II that best reproduces the 
observed abundance pattern among heavy elements in the solar neighborhood 
as well as in the Large and Small Magellanic Clouds.
Given the IMF, the total number of SNe Ia and II from a {\it star particle}
is easily estimated, and this number can then be related to the mass of
ejected heavy elements from a {\it star particle} by making use of Table 2 
of Tsujimoto et al. (1996).
We note that our analysis in this paper is restricted to the metal enrichment
by SNe II only, and the full analysis including the contribution from SNe Ia
will be given elsewhere.

We compute the evolution of spectral energy distribution (SED) of a 
{\it star particle} based on the method of stellar population synthesis
which utilizes the stellar evolutionary tracks for various masses and 
metallicities of stars (Arimoto \& Yoshii 1986).
Using the updated version by Kodama \& Arimoto (1997), we calculate the 
SED for $\lambda=300$\AA$-40000$\AA as a function of elapsed time from
the formation of each {\it star particle}. 
The SED of a whole galaxy is then obtained by summing up SED's of all
{\it star particles} ever formed at different times with different 
metallicities. In this paper the response functions for the $UBVRI$ passbands
are taken from Bessell (1990) and those for the  $JHKL$ magnitudes from 
Bessell \& Brett (1988).

\subsection{Virialized protogalaxy}

Following a standard scenario of the CDM universe ($\Omega_0=1$, 
$H_0=50$ km$\,$s$^{-1}$Mpc$^{-1}$), we consider a protogalaxy as 
originated from the CDM density fluctuation with $\delta M/M$ equal 
to $\nu$ times the {\it rms} value $\sigma$. The fluctuation is normalized 
to unity for a spherical top-hat window of comoving radius 16 Mpc with a bias 
parameter of $b=1$. We assume that this protogalaxy is composed of 
10 \% baryon and 90 \% dark matter in mass and is initially in virial 
equilibrium.

A protogalaxy with the total mass $M$ is assumed to have the density profile
of a singular isothermal sphere with the truncation radius equal to 
$R_{\rm vir}$. The locus of $R_{\rm vir}$ versus $M$ for a $1\sigma$ density
peak is shown by thin line in Fig. 2 (cf. Bardeen et al. 1986).
The local dynamical time of the singular isothermal sphere at the radius of 
$r$ is given by 
\begin{eqnarray}
\tau_{\rm dyn}(r) = 
      \left( \frac{3 \pi^2 R_{\rm vir}}{8 G M} \right)^{\frac{1}{2}} r,
\end{eqnarray}
and the local gas cooling time is given by
\begin{eqnarray}
\tau_{\rm cool}(r) =  \frac{2 \pi G m_{\rm H}^2}{(1-Y)^2 F}
                      \frac{1 }{\Lambda} r^2,
\end{eqnarray}
where $F (=0.1)$ is the baryonic fraction.
Equating $\tau_{\rm dyn}(r)$ to $\tau_{\rm cool}(r)$, we define the cooling 
radius as 
\begin{eqnarray}
R_{\rm cool}(M) = \left( \frac{3 R_{\rm vir}}{32} \right)^{\frac{1}{2}}
   \frac{(1-Y)^2 F \Lambda}{m_{\rm H}^2 G^{\frac{3}{2}}} M^{-\frac{1}{2}}.
\end{eqnarray}
The $R_{\rm cool}$ versus $M$ relation for the solar-abundance gas is shown 
by the upper thick line and the relation for the primordial gas is shown by 
the lower thick line in Fig. 2. 
In a region above (below) the locus of $R_{\rm cool}(M)$, the
cooling time is longer (shorter) than the dynamical time.
For a less massive protogalaxy with $M< 7\times10^{10}M_{\odot}$, the cooling 
is efficient over a whole range of $r$, so that these galaxies can cool and
condense leading to the burst of star formation.
On the other hand, for $M>7\times10^{10}M_{\odot}$, 
the star formation occurs only in the cooling region inside the radius
of $R_{\rm cool}$ $(< R_{\rm vir})$. In a particular case of 
$M=2\times10^{12}M_{\odot}$, the cooling region inside $R_{\rm cool}$ 
contains about 5\% of the total mass.

It is clear from Fig. 2 that the cooling is more efficient for higher 
metallicity. Since synthesized heavy elements by SNe II spread out due to 
the feedback process and stellar motions, the cooling region in which star
formation occurs expands towards an outer part of the system.
Accordingly this expansion of cooling region induces the formation of
massive ellipticals even from the initial condition of low densities like 
$1\sigma$-peaks.

\section{SIMULATIONS}

The formation of the CDM halos through the hierarchical clustering 
has been investigated with $N$-body simulations. 
Dubinski \& Carlberg (1991) argued that their equilibrium density profile
is fitted by $\rho \propto r^{-1}(r+a)^{-3}$ or Hernquist's (1990) profile 
which has a central cusp and resembles de Vaucouleurs' $r^{1/4}$-profile in
projection.
Navarro, Frenk \& White (1997), however, pointed out that their structure
can be approximated as $\rho(r) \propto r^{-1} (r+a)^{-2}$, which is more
extended than the Hernquist profile. 
Fukushige \& Makino (1997) used a high-resolution simulation and
showed that the CDM halos have a steeper central cusp than quoted above.

The resulting structure of the CDM halos depends on the number of particles 
used in the simulation. Besides, there is no definitive view about whether 
the baryonic component has the same equilibrium profile as the dark matter 
halo. Thus, in this paper, we assumed that both baryon and dark matter 
initially have the King profile with the central concentration index of 
$c=2$ (Fig. 3; cf. Binney \& Tremaine 1987).
Since our prime motivation is to clarify the effect of energy feedback in 
the evolution of galaxies, we simply neglect the possible mass-dependent 
profile of dark halos (Navarro, Frenk \& White 1997) and their angular 
momentum distribution (Mao \& Mo 1998). 
This two-component system is made to settle in a virial equilibrium 
from which the gas temperature and the velocity dispersion of dark matter 
are estimated.
Our simulation uses $3 \times 10^4$ gas particles and the same number of 
dark matter particles to set up an initial condition.
As the number of {\it star particles} increases due to star formation, 
the total number of particles increases up to about $10^5$ particles in 
the end of our simulation.

\subsection{Dwarf elliptical galaxies}

We consider a less massive protogalaxy having a total mass of 
$10^{10}M_{\odot}$ with a baryon to dark matter ratio of 1/9.  
The tidal radius of the King profile is 8.45 kpc.
The mean density of the total system is $2.7 \times 10^{-25}$ g cm$^{-3}$,
the mean temperature of the gas is $ 10^{5.1}$ K, the mean velocity 
dispersion of dark matter is 72 km s$^{-1}$, the mean dynamical time is 
$1.3 \times 10^8$ yrs, and the mean cooling time is $7.6 \times 10^7$ yrs.
The gravitational softening parameter is adopted as 0.02 kpc for gas 
particles, 0.05 kpc for dark matter, and 0.03 kpc for {\it star particles}.

Owing to the efficient radiative cooling mainly through collisional 
excitation of H and He${^+}$, the gas temperature rapidly drops,
which induces a dynamical contraction of gas and dark matter by the 
self-gravity.
When the gas temperature becomes close to $10^4$ K and stops decreasing, 
a quasi-isothermal contraction is established.
The density in the central region increases by the accretion of the 
surrounding gas, and eventually the intensive star formation is triggered.
When massive stars begin to explode as SNe II, the gas in the vicinity of
SNe II acquires the thermal energy and synthesized heavy elements released 
from SNe II. Then, the gas temperature locally increases up to about 
$10^{7.5}$K, and subsequent formation of stars is virtually halted. 
About 5\% of the initial gas mass is used up in this formation of the 
first generation stars.  

Figure 4 shows the snapshots for the projected particle positions and the 
integrated energy spectra as a function of elapsed time from 
$0.5 \times 10^7$ to $1.0 \times 10^9$ yrs.
The top three panels in each column show the spatial distributions of dark 
matter, gas, and stars, respectively, projected onto the $x-y$ plane.
The bottom panel shows the spectral energy distribution (SED) from stellar 
populations in the evolving galaxy.
The supernova-driven gas flow is generated outwards from the center of the 
protogalaxy.
This outflow collides with the inflow of the gas accreting from outside,
and the high-density super-shell is eventually formed.  
Figure 5 shows the radial profiles of the gas density (top panel), 
the temperature (middle panel), and the radial velocity (bottom panel) at 
the elapsed time of $ 1.0 \times 10^7 $ yrs. 
We find from this figure that the shock waves propagate outwards with the
shock front at $\sim 0.5$ kpc and the hot cavity is created inside the 
super-shell. 

While the gas is continuously swept up by the super-shell, the gas density 
further increases due to the enhanced cooling rate in the already dense 
shell. Then the secondary star formation begins within the super-shell,
and subsequent SNe further accelerate the outward expansion of 
the shell. This situation is clearly understood from the second panels of
Fig. 4 in a time sequence from $1.0 \times 10^7$ to $3.0 \times 10^7$ yrs.
Star formation in the expanding shell continue for $\sim 5.0 \times 10^7$ 
yrs until the gas density in the shell becomes too low to
form new stars.  About 20\% of the initial gas mass is turned into stars
in this stage.  
Finally, the outflowing gas escapes from the gravitational potential of the 
whole system and is ejected into the intergalactic space. The final stellar 
system possesses $\sim 25 \%$ of the initial gas mass.
    
The stars initially have the velocity vectors of the gas from which 
they were formed.  Therefore, the first generation stars have zero 
systematic velocity, but the later generation stars have a large outward 
radial velocity component. The oscillation of swelling and contraction of 
the stellar system continues for several $10^8$ yrs, and the system becomes
settled in a quasi-steady state until $\sim 1 $ Gyrs.  
Consequently the system has a large velocity dispersion and a large core.  
The surface mass distribution is approximately exponential and is more 
extended than de Vaucouleurs' $r^{1/4}$-profile.

Stars are formed for the most part before the gas is fully polluted to 
the yield value of the synthesized heavy elements.  
The average metallicity of the stars in the system is as low as 
[Fe/H] $\sim -2.4$.  This metallicity is consistent with a range covered 
by the observations, but is much lower than those of normal galaxies 
(Dekel \& Silk 1986; Yoshii \& Arimoto 1987). 
One outstanding feature discovered by our simulation is a {\it positive} 
metallicity gradient in this system which is in sharp contrast to the 
observed negative gradient for massive galaxies 
(Carollo, Danziger \& Buson 1993).  
The star-forming site moves outwards with the expanding shell and the gas 
in this shell is gradually enriched with synthesized heavy elements from 
SNe II.  
Stars of later generations are necessarily born at larger radii with larger 
metallicities, leading to the emergence of a positive metallicity gradient 
in the resulting stellar system.
Figure 6 shows the projected distribution of stellar metallicity at different 
elapsed times of $1.0 \times 10^7, 2.0 \times 10^7, 3.0 \times 10^7$, 
and $1.0 \times 10^9$ yrs. The distribution does not change with time and the 
system should keep such a metallicity gradient during the age of the universe.

Figure 7 shows the projected surface brightness distribution at 15 Gyrs
in the {\it B, V, R, I} and {\it K} bands. These profiles are plotted 
against a linear scale of the radius in kpc or a quatic root of the radius 
in kpc. The resulting system is characterized by an exponential profile 
rather than de Vaucouleurs' $r^{1/4}$-profile.
Figure 8 shows the projected color profiles 
at 15 Gyrs for {\it V--K} (circles), {\it B--R} (squares), {\it B--V}
(diamonds), and {\it V--R} (triangles). These profiles are scaled vertically
in this figure to coincide with each other at the galaxy center.
The result is consistent with the observed trend of redder colors at larger
radii for dwarf galaxies (Vader {\it et al.} 1988; 
Kormendy \& Djorgovski 1989; Chaboyer 1994).  

In our simulation of dwarf ellipticals, the total stellar mass is
smaller when a flatter IMF is adopted. However, the stellar metallicity
is not so sensitive to the IMF slope and remains about one tenth of
than the solar in agreement with observations of nearby dwarf galaxies. 
This is because for the flatter IMF the number of supernovae is larger so 
that the heated gas and heavy elements therein are blown out of the system 
at a larger rate. 

We also examined how our simulation depends on the SFR coefficient $C$ in 
equation 12. 
For $C=0.1$ and $C=1.0$, we carried out the simulations with the same 
initial condition.
Our simulation run gives the total stellar mass of $3.12 \times 10^8 M_\odot$ 
for $C=0.1$ and $4.27 \times 10^8 M_\odot$ for $C=1.0$,
indicating that the result is insensitive to the adopted value of $C$ as 
noted by Katz (1992) and Katz, Weinberg \& Hernquist (1995).

\subsection{Normal elliptical galaxies}

In this section we study the formation of a more massive system
for the purpose of comparison with a less massive system of dwarf elliptical
galaxies. We consider a protogalaxy having a total mass of 
$10^{12}M_{\odot}$ with a baryon to dark matter ratio of 1/9. 
The tidal radius of the King profile is 82.8 kpc.
The mean density of the total system is $2.8 \times 10^{-26} $ g cm$^{-3}$,
the mean temperature of the gas is $ 10^{6.1}$ K, and the mean velocity 
dispersion of dark matter is 228 km s$^{-1}$, the mean dynamical time 
is $4.0 \times 10^8$ yrs, and the mean cooling time is $5.7 \times 10^{10}$ 
yrs.
The gravitational softening parameter is adopted as 0.2 kpc for gas 
particles, 0.5 kpc for dark matter, and 0.3 kpc for {\it star particles}.

Figure 9 shows the snapshots for the projected particle positions and the 
integrated energy spectra as a function of elapsed time from 
$0.1 \times 10^9$ to $2.0 \times 10^9$ yrs.
The top three panels in each column show the spatial distributions of dark 
matter, gas, and stars, respectively, projected onto the $x-y$ plane. 
The bottom panel shows the SED from the evolving galaxy.

Contrary to dwarf ellipticals, a massive protogalaxy is considered to evolve 
from relatively low gas density and high virial temperature, so that the gas
does not cool rapidly. The protogalaxy initially in virial 
equilibrium shrinks quasi statically, and the gas density near the galaxy 
center gradually increases. When the gas density rises sufficiently, 
the temperature suddenly drops due to the thermal instability.
Figure 10 shows that the gas temperature drops down to about 
$10^4$ K and the gas density further increases in the central region 
where active star formation starts to occur.

Newly born massive stars heat up the surrounding gas and at the same
time the rate of their formation is suppressed in the heated gas.
Therefore, higher or lower rate of star formation is stabilized and 
self-regulated when all star-forming activities are confined
within the deep gravitational potential of a massive galaxy.

Figure 11 shows the projected surface brightness distribution at 15 Gyrs 
plotted against a linear scale of the radius in kpc or a quatic root of 
the radius in kpc.  
The surface brightness distribution at 15 Gyr is similar to an 
$r^{1/4}$-profile, and the integrated blue luminosity is $M_B=-21.5$ mag.

Since massive stars enrich the gas with synthesized heavy elements from which
new stars are subsequently born, the stellar metallicity increases in 
proportion to the heavy element abundances in the gas.
In the central region of the galaxy, cycles of stellar birth and death 
continue until the metallicity approaches the yield value, whereas in the
outer part of the galaxy the gas is used up in star formation before it is
significantly polluted.
Consequently as shown in Fig. 12, there appears a negative gradient of 
stellar metallicity across the system corresponding to bluer colors at 
larger radii as observed in normal galaxies 
(e.g. Carollo, Danziger \& Buson 1993).

We however note that the average stellar metallicity is [Fe/H]$\sim -0.2$,
which is about three times lower than the observed metallicity of [Fe/H] 
$\sim +0.3$, either because Salpeter's IMF adopted in our simulation is too 
steep or because an initial density of the protogalaxy taken from a 
$1\sigma$ density peak is too low to form massive elliptical galaxies.
Our numerical experiments indicate that a flatter IMF with $x=1.05$ is able 
to reproduce the observed metallicity. In order to refine the
model, we need more extensive simulations from different initial conditions
by including the iron supply from SNe Ia in the gas. Comprehensive analysis 
along this line will be reported elsewhere.

\section{SUMMARY AND CONCLUSION}

We have developed a three-dimensional $N$-body/SPH simulation code combined 
with stellar population synthesis and applied it to the study of dynamical,
chemical and spectro-photometric evolution of dwarf and normal elliptical 
galaxies embedded in a dark matter halo. A protogalaxy is assumed to be a 
virialized non-rotating sphere in a $1\sigma$ CDM perturbation enclosing the 
total mass of $10^{10}M_\odot$ and $10^{12}M_\odot$ with 10 \% baryonic mass.

For dwarf ellipticals, the gas temperature rapidly decreases by the 
efficient cooling and the protogalaxy starts to collapse due to the 
self-gravity of gas and dark matter. When the gas density near the galaxy
center becomes large, star formation burst takes place 
(SFR $\sim 23 M_\odot$ yr$^{-1}$) and the collapse is halted by 
supernova-driven wind from the central star-forming region. 
About 75 \% of the initial gas mass is then lost from the system on a much 
shorter time scale compared to the mean crossing time of a star in the system. 
Thereby the system expands and recovers a new equilibrium state as a 
loosely-bound stellar system exhibiting an exponential structure.

On the other hand, normal ellipticals are considered to evolve from 
ten times lower gas density and ten times higher virial temperature.
The gas is thermally stable having a relatively longer cooling time.
In our simulation, most of the gas is used up in formation of stars and
only a small fraction of gas is evaporated out of the system in the form
of irregular gaseous blobs. 
Since the time scale of this gas removal is much longer than the crossing 
time of a star, the stellar system remains in quasi-equilibrium and evolves 
into an $r^{1/4}$-structure as observed.

The dynamical evolution of the system is determined not only by the amount 
of gas removal but also by its time scale. 
These decisive factors of gas removal depend critically on the cooling 
efficiency which is systematically higher for less massive protogalaxies
in the CDM universe (see Fig. 2). In particular their efficient cooling 
causes a significant dynamical impact in making an exponential structure of 
dwarf galaxies. However, in a low-mass end below $10^8 M_\odot$, protogalaxies
have much more efficient cooling, so that a large amount of gas is locked up
in formation of the first-generation stars until they start to explode as 
supernovae. Consequently supernovae no longer affect the subsequent evolution 
of the stellar system and very low-mass galaxies could survive as first 
demonstrated by Yoshii \& Arimoto (1987).

It is therefore intriguing to simulate spheroidal stellar systems along
their mass sequence from giant elliptical galaxies to compact globular 
clusters in order to see whether the energy feedback works to suppress the
formation of dwarf galaxies only and modifies the shape of the luminosity
function as observed.

\bigskip

We are grateful to T. Shigeyama, M. Chiba and T. Tsujimoto for many 
fruitful discussions, and to T. Kodama for providing us the tables of 
population synthesis. 
This work has been supported in part by the Research Fellowship of the 
Japan Society for the Promotion of Science for Young Scientists (6867),
the Grant-in-Aid for Scientific Research (05242102, 06233101), and 
the Center-of-Excellence Research (07CE2002) of the Ministry of Education, 
Science, Sports, and Culture in Japan.
The computations were mainly carried out on the ``Remote-GRAPE'' system at 
University of Tokyo and partly done on Fujitsu VPP-300 at the National 
Astronomical Observatory in Japan, and Fujitsu VPP-500 at the Institute 
of Physical and Chemical Research (RIKEN).

\clearpage

\figcaption[metal_cooling.ps]{
The cooling rate $\Lambda$ divided by the square of number density $n$ 
of gas particles is plotted against the temperature.
Each curve shows the cooling rate for the value of [Fe/H] indicated.
The bottom curve corresponds to the zero metallicity.
}

\figcaption[diagram_1s.ps]{
Radius of a protogalaxy as a sequence of the total mass with a baryon
to dark matter ratio equal to 1/9. Thick lines show the cooling radius
$R_{\rm cool}$ from $\tau_{\rm dyn}=\tau_{\rm cool}$ for the solar 
metallicity (upper line) and the zero metallicity (lower line).
Thin line shows the virial radius $R_{\rm vir}$ from a $1\sigma$ density
perturbation in the standard CDM universe ($\Omega_0=1$, 
$H_0=50$ km$\,$s$^{-1}$Mpc$^{-1}$).
}

\figcaption[king_profile.ps]{
Initial density profile of dark matter having the mass of 
$9 \times 10^{9} M_\odot$.
The solid line indicates the exact density distribution of the King profile 
with the central concentration index of $c=2$, and the crosses are the 
initial densities set up in our simulation. 
The inset shows the projected particle distribution in the $x-y$ plane.
}

\figcaption[dwarf_snap.ps]{
Snapshots for the projected particle positions and the spectral energy
distribution (SED) at five different elapsed times for our simulation 
run of a protogalaxy with the total mass of $10^{10}M_{\odot}$.
The top three panels in each column show the spatial distributions  of 
dark matter, gas, and stars, respectively, projected onto the $x-y$ plane.
The bottom panel shows the SED from stellar populations in the evolving 
galaxy.
}

\figcaption[dwarf_dens_temp_vel.ps]{
Snapshot for the protogalaxy with the total mass of $10^{10}M_{\odot}$
at the elapsed time of $1.0 \times 10^7$ yrs. 
Shown are the radial profile of the gas density (top panel),
the gas temperature (middle panel), and the radial component of the gas 
velocity (bottom panel).
}

\figcaption[surf_evl_metal.ps]{
Projected distributions of stellar metallicity [Fe/H] at four different 
elapsed times for our simulation run of a protogalaxy with the total 
mass of $10^{10}M_\odot$.
}

\figcaption[dwarf_surf_brightness.ps]{
Projected distribution of surface brightness at 15 Gyrs for our simulation
 run of a protogalaxy with the total mass of $10^{10}M_{\odot}$. 
Thick lines from top to bottom are the surface brightnesses in the 
{\it B, V, R, I} and {\it K} bands, respectively, plotted against a liner 
scale of the radius in kpc or a quatic root of the radius in kpc. 
A mean $B$-band distribution observed for a majority of dwarf galaxies is 
also shown for reference (e.g. Patterson \& Thuan 1996).
}

\figcaption[dwarf_delta_color_metal.ps]{
Projected distribution of integrated color at 15 Gyrs for our simulation
run of a protogalaxy with the total mass of $10^{10}M_{\odot}$. Shown are
the colors of {\it V--K} (circles), {\it B--R} (squares), {\it B--V} 
(diamonds), and {\it V--R} (triangles). Their values are offset at the 
galaxy center. The inset shows the projected distribution of stellar 
metallicity at 15 Gyrs}

\figcaption[giant_snap.ps]{
Snapshots for the projected particle positions and the spectral energy
distribution (SED) at five different elapsed times for our simulation 
run of a protogalaxy with the total mass of $10^{12}M_{\odot}$.
The top three panels in each column show the spatial distributions  of 
dark matter, gas, and stars, respectively, projected onto the $x-y$ plane.
The bottom panel shows the SED from stellar populations in the evolving 
galaxy.
}

\figcaption[giant_dens_temp_vel.ps]{
Snapshot for the protogalaxy with the total mass of $10^{12}M_{\odot}$
at the elapsed time of $0.5 \times 10^9$ yrs. 
Shown are the radial profile of the gas density (top panel),
the gas temperature (middle panel), and the radial component of the gas 
velocity (bottom panel).
}

\figcaption[giant_surf_brightness.ps]{
Projected distribution of surface brightness at 15 Gyrs for our simulation
run of a protogalaxy with the total mass of $10^{12}M_{\odot}$. Thick lines
from top to bottom are the surface brightnesses in the {\it B, V, R, I} and 
{\it K} bands, respectively, plotted against a liner scale of the radius 
in kpc or a quatic root of the radius in kpc. 
}

\figcaption[giant_delta_color_metal.ps]{
Projected distribution of integrated color at 15 Gyrs for our simulation
run of a protogalaxy with the total mass of $10^{12}M_{\odot}$. Shown are
the colors of {\it V--K} (circles), {\it B--R} (squares), {\it B--V} 
(diamonds), and {\it V--R} (triangles). Their values are offset at the 
galaxy center. The inset shows the projected distribution of stellar 
metallicity at 15 Gyrs.
}

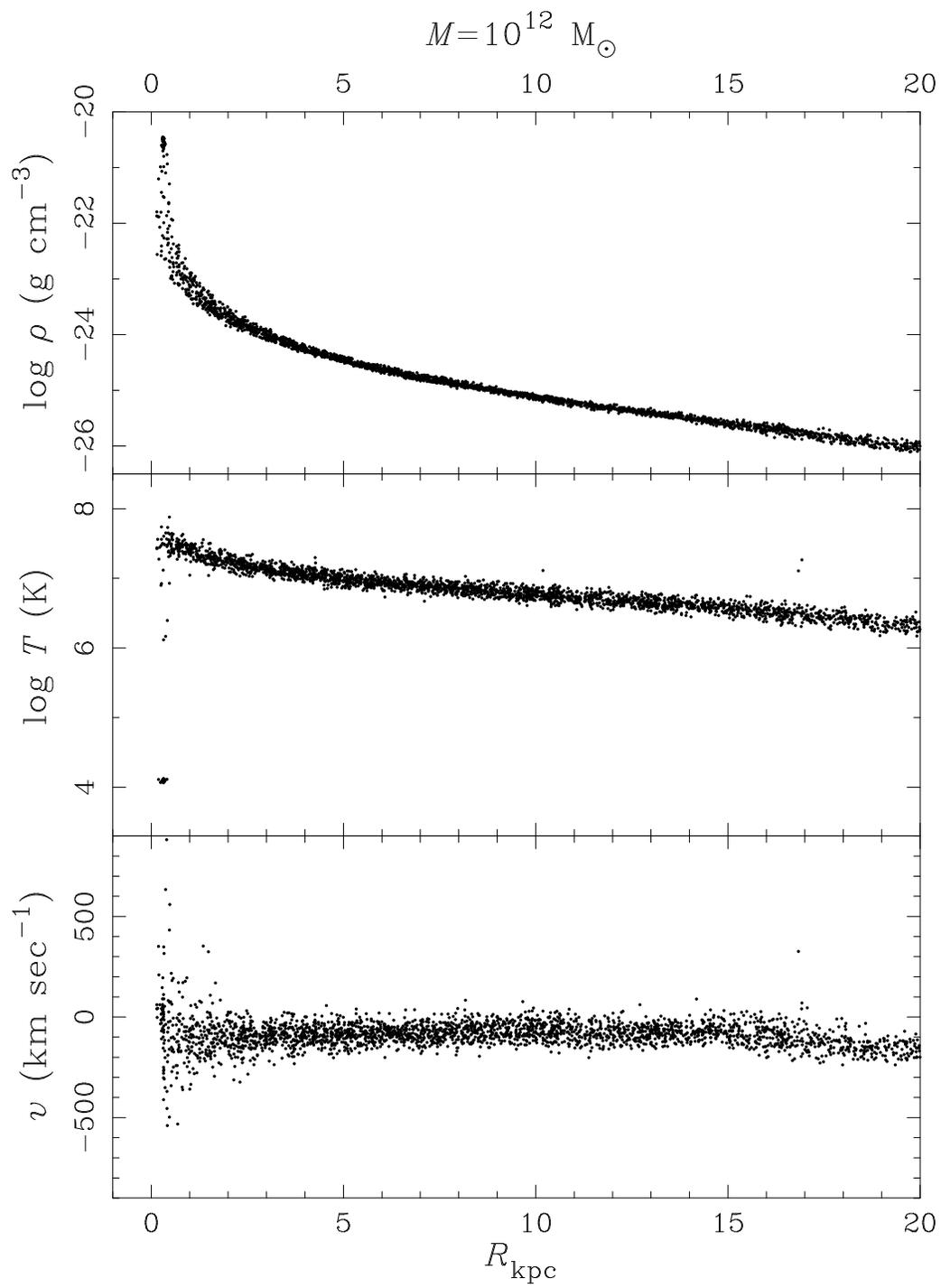
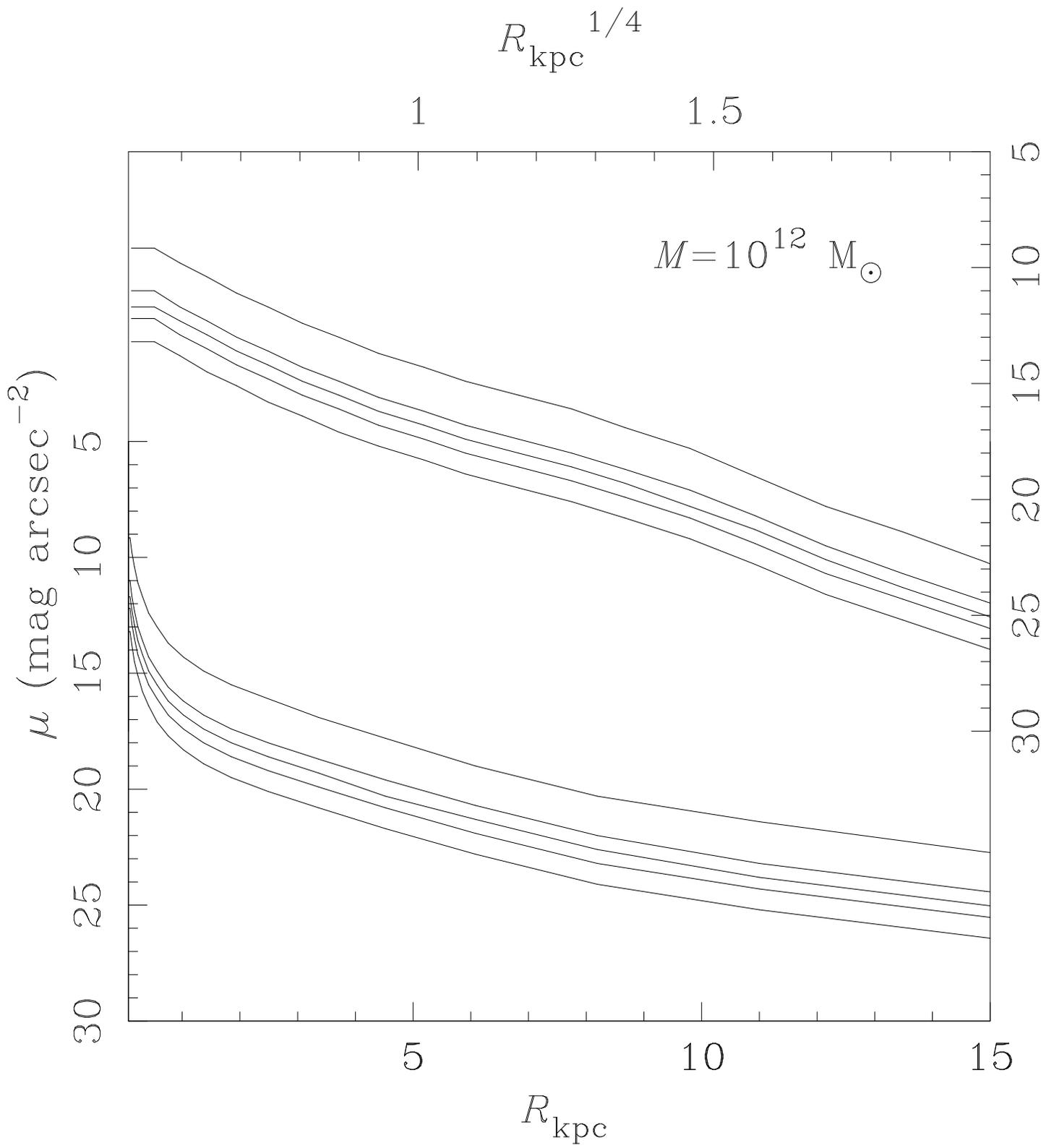
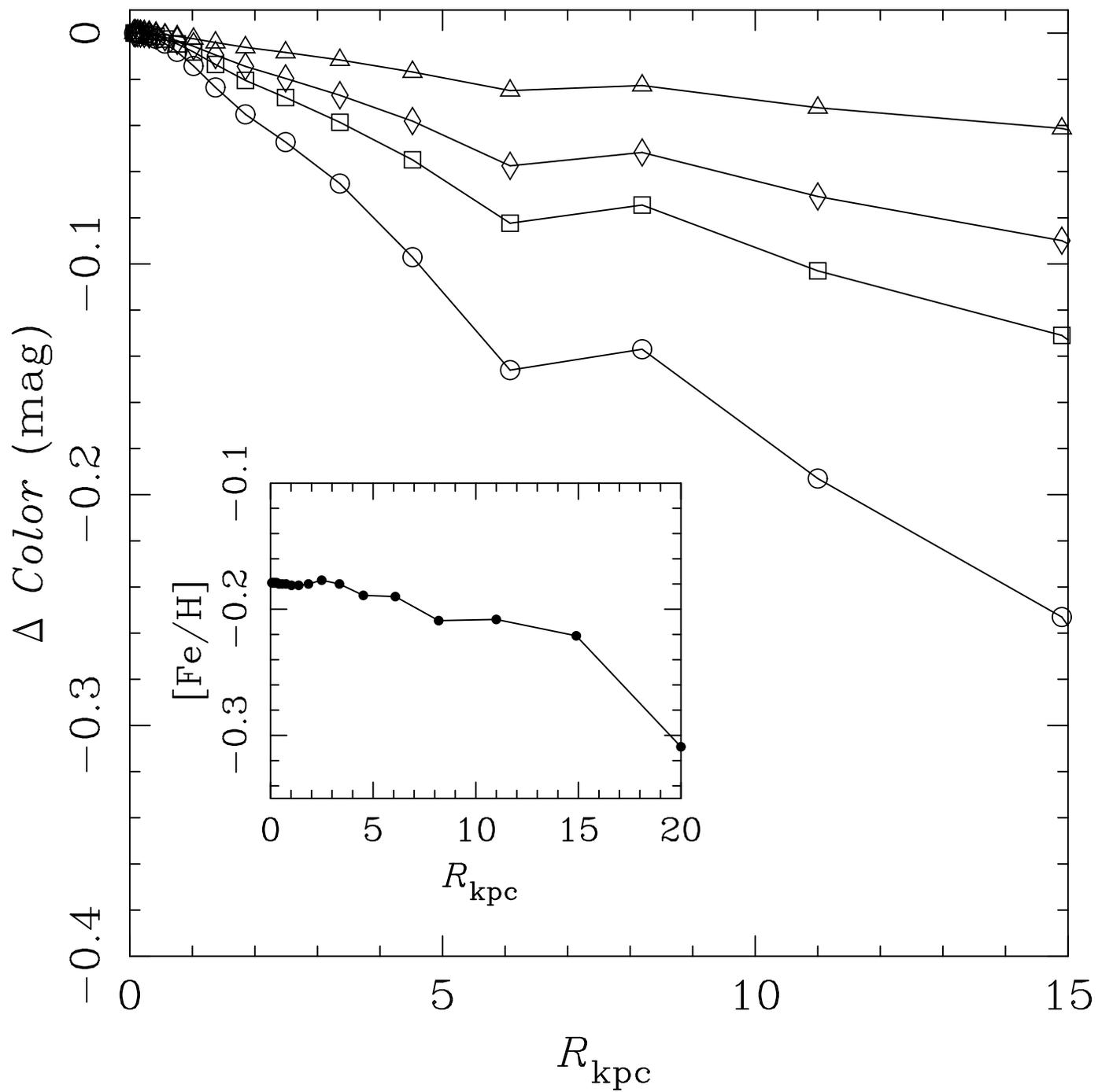
\end{document}